\pdfoutput=1
\documentclass{ar-1col}

\usepackage[numbers]{natbib}

\newcommand{\pstar}{$p^{\star}$}
\newcommand{\xstar}{$x^{\star}$}
\newcommand{\Tstar}{$T^{\star}$}
\newcommand{\Tc}{$T_{\rm c}$}
\newcommand{\Hc}{$H_{\rm c2}$}
\newcommand{\Hvs}{$H_{\rm vs}$}

\newcommand{\RH}{$R_{\rm H}$}
\newcommand{\nH}{$n_{\rm H}$}

\newcommand{\mstar}{$m^{\star}$}

\newcommand{\pFS}{$p_{\rm FS}$}
\newcommand{\pcdw}{$p_{\rm CDW}$}

\newcommand{\Cel}{$C_{\rm el}$}
\newcommand{\CelT}{$C_{\rm el}/T$}

\jname{Xxxx. Xxx. Xxx. Xxx.}
\jvol{10}
\jyear{2018}
\doi{10.1146/((please add article doi))}

\begin{document}
\pdfpagewidth=8.5in
\pdfpageheight=11in

\markboth{Proust \& Taillefer}{The remarkable
underlying ground states of cuprate superconductors}

\title{The remarkable underlying
ground states of cuprate superconductors}

\author{
Cyril Proust$^{1,2}$ and Louis Taillefer$^{2,3}$
\affil{$^1$Laboratoire National des Champs Magn\'{e}tiques Intenses (CNRS, EMFL, INSA, UJF, UPS), Toulouse 31400, France;
email: cyril.proust@lncmi.cnrs.fr}
\affil{$^2$Canadian Institute for Advanced Research, Toronto, Ontario M5G 1Z8, Canada}
\affil{$^3$Institut quantique, D\'{e}partement de physique  \&  RQMP, Universit\'{e} de Sherbrooke, Sherbrooke,  Qu\'{e}bec J1K 2R1, Canada;
email: louis.taillefer@usherbrooke.ca}
}

\begin{abstract}
Cuprates exhibit exceptionally strong superconductivity.
To understand why, it is essential to elucidate the nature of the electronic
interactions that cause pairing.
Superconductivity occurs on the backdrop of several
underlying electronic phases,
including a doped Mott insulator at low doping, a strange metal at high doping,
and an enigmatic pseudogap phase in between -- inside which a phase of charge-density-wave order
appears.
In this Article, we aim to shed light on the nature of these remarkable
phases
by focusing on the limit as $T \to 0$, where experimental signatures
and theoretical statements become sharper.
We therefore survey the ground state properties of cuprates
once superconductivity has been removed
by the application of a
magnetic field,
and distill their key universal features.

\end{abstract}

\maketitle


\begin{figure}[t]
\includegraphics[width=6.3in]{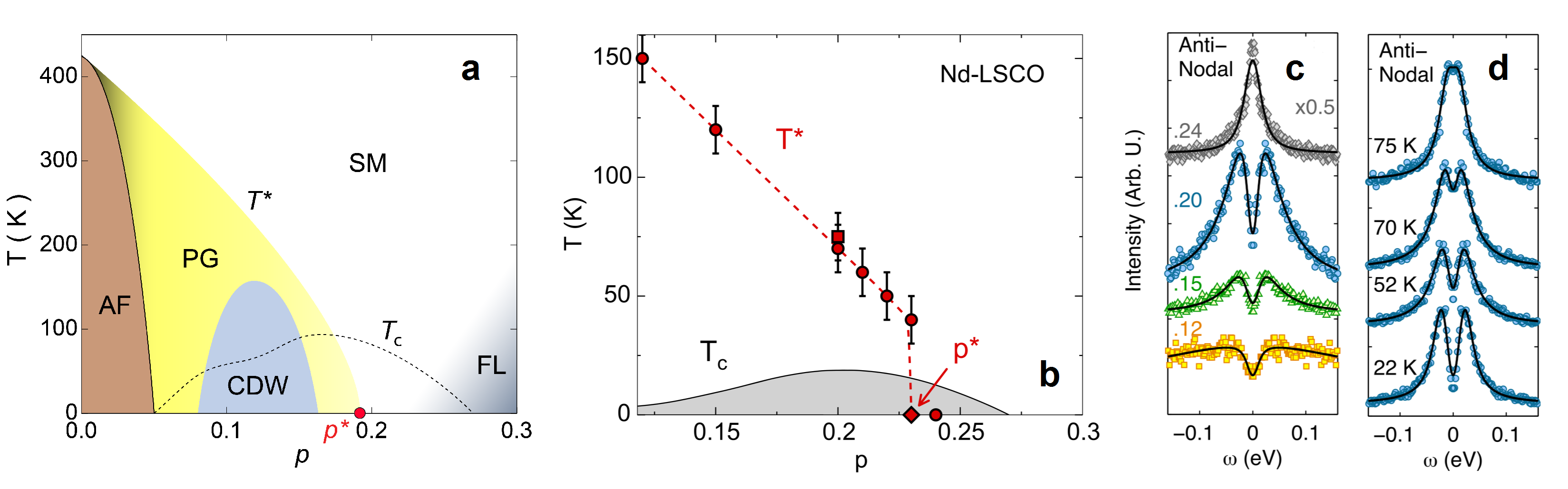}
\caption{
{\bf Phase diagram of hole-doped cuprates.}
{\bf a)}
In zero field, superconductivity exists in a dome below \Tc~(dashed line).
When it is removed by a magnetic field,
various underlying ground states are revealed:
1) Doped Mott insulator with antiferromagnetic order, on the far left (brown, AF);
2) Pseudogap (PG) phase below a temperature \Tstar~(yellow, PG), ending at a $T=0$ critical point \pstar~(red dot);
3) Charge-density-wave phase
(blue, CDW),
contained inside the pseudogap phase;
4) a strange metal just above \pstar~(white region),
which gives way to a Fermi liquid at highest doping~(grey region).
{\bf b)}
Phase diagram of Nd-LSCO, with the pseudogap temperature \Tstar~measured by resistivity (circles)
and ARPES (square; panels c, d), ending at the critical point \pstar~(from ref.~\cite{Collignon2017}).
{\bf c)}
ARPES spectra showing the pseudogap in Nd-LSCO measured just above \Tc~at four dopings, as indicated \cite{Matt2015}.
The pseudogap is seen to close between $p = 0.20$ and $p = 0.24$, consistent with \pstar~$=0.23$.
{\bf d)}
ARPES spectra at $p=0.20$ vs temperature \cite{Matt2015}.
The pseudogap
is seen to close
at \Tstar~$=75$~K (square in panel b).
}
\label{PhaseDiagram}
\end{figure}

\section{INTRODUCTION}

After more than three decades, cuprates continue to fascinate physicists because
of a persistent sense -- a growing conviction -- that these materials host novel quantum phenomena.
And these arise from electron interactions that are most likely
also
responsible for the exceptionally strong superconductivity.

The repulsive interaction between electrons in cuprates is so strong that when there is one electron on every Cu site
of their CuO$_2$ planes, a Mott insulator forms in which no motion is possible.
By removing electrons, or adding $p$ holes (per Cu site), electron motion is restored,
and at high enough $p$ cuprates become well-behaved metals.
The unusual phenomena occur in the intermediate regime, between the Mott insulator at $p=0$
and the Fermi liquid at $p>0.3$ (Fig.~\ref{PhaseDiagram}a).

This is where superconductivity lives, below a critical temperature \Tc~that forms a dome (Fig.~\ref{PhaseDiagram}a),
peaking at a value that can exceed 150~K -- halfway to room temperature.
In this Article, we ask the following question:
How does the underlying normal state -- from which superconductivity emerges -- evolve with doping?
In particular, we focus on the ground state, as $T \to 0$, accessed by suppressing superconductivity
with a large magnetic field.

At $T=0$, in the absence of superconductivity, the key event on the path from Fermi liquid to Mott insulator
is the onset of the pseudogap phase, at a critical doping \pstar~(red dot in Fig.~\ref{PhaseDiagram}).
One of the most remarkable -- and puzzling -- phenomena in condensed-matter physics, the pseudogap phase
exists in all hole-doped cuprates below a temperature \Tstar~that decreases with doping to end at \pstar~(Fig.~\ref{PhaseDiagram}).
We will discuss what high-field studies reveal about the ground state of cuprates, both inside ($p <$~\pstar)
and outside ($p >$~\pstar) the pseudogap phase.
The latter region presents another major puzzle of condensed-matter physics:
a perfectly $T$-linear dependence of resistivity as $T \to 0$.
While not unique to cuprates, this is where
this remarkable
phenomenon is strongest.

Before we begin, it is important to mention that the phase diagram of Fig.~\ref{PhaseDiagram},
with its \Tc~dome straddling a critical point,
is reminiscent of that found in several families of materials, all part of the same general paradigm --
the paradigm of an antiferromagnetic quantum critical point (QCP)
\cite{Monthoux2007,Taillefer2010}.
These include heavy-fermion metals, iron-based superconductors and quasi-1D organic
conductors.
The latter are a good archetype, because of their simple Fermi surface,
and their key properties include :

\begin{enumerate}
\item AF phase ends at a QCP, located at $X$*
\item $d$-wave superconductivity forms a dome surrounding $X$*
\item Fermi surface is reconstructed (by AF order) below $X$*
\item $T$-linear resistivity at $X=$~$X$*
\item Fermi-liquid $T^2$ resistivity at $X\gg$~$X$*
\end{enumerate}

where $X$ is the tuning parameter, {\it e.g.} pressure.
In materials like (TMTSF)$_2$PF$_6$, there is little doubt
that AF spin fluctuations, measured by NMR, are responsible for $d$-wave pairing and $T$-linear
scattering \cite{Lebed2008}.

When cuprates are doped with electrons rather than holes,
their properties are also consistent with the AF QCP paradigm (see Sidebar).
In what ways, then, are hole-doped cuprates different?
First, a quantitive difference:
their electron interactions are stronger, as measured by their higher \Tc~and their higher effective mass \mstar, for example.
Secondly, a qualitative difference:
they have a pseudogap phase, for which there is no real equivalent in electron-doped cuprates
\cite{Armitage2010}.

\begin{marginnote}[]
\entry{AF}{antiferromagnetic}
\entry{PG}{pseudogap}
\entry{CDW}{charge density wave}
\entry{SDW}{spin density wave}
\entry{\Tc}{superconducting transition temperature in zero field}
\entry{\Tstar}{pseudogap temperature}
\entry{\pstar}{pseudogap critical point}
\entry{QCP}{quantum critical point}
\entry{NCCO}{Nd$_{2-x}$Ce$_x$CuO$_4$}
\entry{PCCO}{Pr$_{2-x}$Ce$_x$CuO$_4$}
\entry{LCCO}{La$_{2-x}$Ce$_x$CuO$_4$}

\end{marginnote}

\begin{textbox}[t]\section{ELECTRON-DOPED CUPRATES}
Electron-doped cuprates are consistent with the paradigm of an AF QCP -- with AF order,
FS reconstruction
and $T$-linear resistivity all organized around a quantum critical point at \xstar, in NCCO, PCCO and LCCO
\cite{Armitage2010}.
With decreasing
electron doping
$x$, at $T=0$, the AF correlation length increases rapidly below \xstar~\cite{Motoyama2007}, the critical doping where the Fermi surface
undergoes a sharp transition from a large cylinder containing $n = 1 - x$ holes,
as seen in \RH~\cite{Dagan2004}, QOs \cite{Helm2009} and ARPES \cite{Armitage2002,Matsui2007},
to small closed pockets (seen in QOs), and eventually to a small anti-nodal electron pocket (seen in ARPES),
consistent with $n \simeq - x$ (from \RH).
ARPES shows that the reconstruction is consistent with an AF Brillouin zone with ordering wavevector $Q = (\pi, \pi)$~\cite{Matsui2007}.
$T$-linear resistivity is observed as $T \to 0$, at and slightly above \xstar~\cite{Fournier1998,Jin2011,Sarkar2017}.
At $x \gg$~\xstar, $\rho \sim T^2$~\cite{Jin2011};
at  $x <$~\xstar, upturns appear in $\rho(T)$ at low $T$~\cite{Sarkar2017,Tafti2014}.
%
Unlike in hole-doped cuprates, the 'pseudogap' observed by ARPES in NCCO is clearly associated with the AF order \cite{Matsui2007}:
as a function of angle around the FS, it is maximal at the 'hot spots' where the FS intersects the AF zone boundary;
as a function of $x$, its spectral weight decreases in tandem with the AF correlation length;
as a function of decreasing $T$, it forms below the same temperature \Tstar~where optics sees a 'pseudogap' open, which
is where the AF correlation length exceeds the thermal de Broglie wavelength \cite{Kyung2004}.
%
\end{textbox}

\section{REMOVING SUPERCONDUCTIVITY}

In this Article, we are interested in the underlying ground state of cuprates in the absence of superconductivity.
In other words, we want to investigate the normal state of electrons at zero temperature.
Experimentally, there are two main ways of removing superconductivity:
applying a large magnetic field and adding impurities like Zn to the sample.
We will focus on the former approach, but occasionally mention the latter.
The first question is: How much field is needed?
Note that the field is much more effective if applied normal to the CuO$_2$ plane (parallel to the $c$ axis),
so throughout this Article $H$ is
applied in that direction.

In cuprates, it is difficult to determine the upper critical field \Hc~needed to suppress superconductivity from electric
or thermo-electric transport measurements.
For example, the electrical resistivity $\rho$ first increases in the flux-flow regime (due to vortex displacement)
and then continues to increase
at high fields, either because of a positive
normal-state
magnetoresistance (MR)
or because of superconducting fluctuations above \Hc~-- difficult to say.
Moreover, sample inhomogeneity can lead to an apparent \Hc~larger than the bulk because small regions with a higher \Hc~show up strongly in the resistivity.
\begin{figure}[t]
\includegraphics[width=5in]{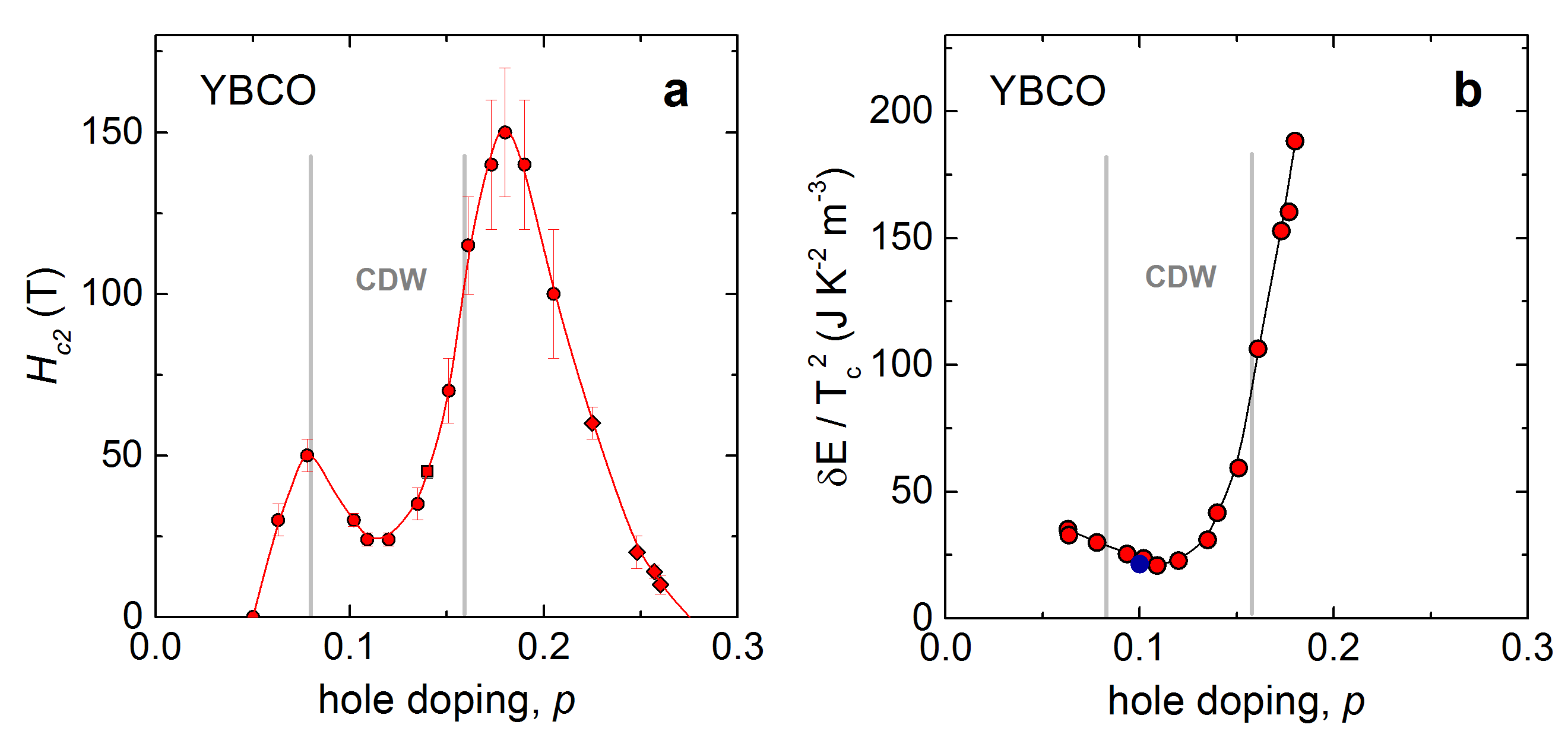}
\caption{
{\bf Critical field and condensation energy.}
{\bf a)}
Doping dependence of the upper critical field at $T \to 0$, \Hc,
in YBCO (circles), Y124 (square) and Tl2201 (diamonds).
(Here, $p$ for Tl2201 is defined via \Tc, as for YBCO.)
For values above 70~T, \Hc~in YBCO is obtained by extrapolating \Hvs~at $T \rightarrow 0$, as are values for Tl2201 (see text).
%
From ref.~\cite{Grissonnanche2014}.
{\bf b)}
Condensation energy $\delta E = H_{\rm c}^2 / 2 \mu_0$,
plotted as $\delta E /$\Tc$^2$ vs $p$ in YBCO (red dots).
From ref.~\cite{Grissonnanche2014}.
The blue dot at $p=0.1$
is the value predicted from the specific heat coefficient $\gamma$ in YBCO \cite{Klein2018}
via
$\delta E = (3 \Delta_0^2 / 8 \pi^2 k_{\rm B}^2) \gamma$, with $\Delta_0 = 2.8~k_{\rm B}$\Tc.
%
%
%
The grey vertical lines mark the boundaries of the CDW phase.
%
%
%
%
%
%
}
\label{Hc2}
\end{figure}


%

\begin{marginnote}[]
\entry{\Hc}{upper critical field}
\entry{QOs}{quantum oscillations}
\entry{ARPES}{angle-resolved photoemission spectroscopy}
\entry{STM}{scanning tunneling microscopy}
\entry{ADMR}{angle-dependent magneto-resistance}
\entry{FS}{Fermi surface}
\entry{YBCO}{YBa$_{2}$Cu$_3$O$_y$}
\entry{Y124}{YBa$_2$Cu$_4$O$_8$}
\entry{Tl2201}{Tl$_2$Ba$_2$CuO$_{6+\delta}$}

\end{marginnote}

One of the few physical properties of cuprates found to exhibit a sharp anomaly at \Hc~is the thermal conductivity $\kappa$.
In clean samples at low temperature ($< 10$~K), the electronic mean free path $l_0$ can be much larger than the inter-vortex separation at $H =$~\Hc.
As a result, when $H$ is decreased below \Hc, the quasiparticle mean free path is suddenly curtailed by vortex scattering, causing a precipitous drop in $\kappa(H)$.
This effect is seen in any clean type-II superconductor.
It was used, for example, to directly measure \Hc~$= 24 \pm 2$~T at $T \to 0$ in YBCO at $p = 0.11-0.12$ \cite{Grissonnanche2014}.
%
This is consistent with recent specific heat \cite{Klein2018} and NMR \cite{ZhouNMR2017} measurements showing that $C(H)$ and the Knight shift
do saturate above 25~T at $T=$~2-3~K.
It was also shown that the field
\Hvs$(T)$ above which $\rho(H)$ becomes non-zero,
at the transition from vortex solid to vortex liquid,
is such that \Hvs$(0) =$~\Hc$(0)$ as $T \to 0$, {\it i.e.}~there is no vortex liquid at $T = 0$ \cite{Grissonnanche2014}.
%
%
Therefore, by measuring \Hvs$(T)$ vs $T$ via the resistivity and extrapolating to $T = 0$ one can estimate \Hc~(at $T = 0$).
Applying this procedure to an extensive set of high-field data on YBCO yields the phase diagram of \Hc~vs $p$
across the full doping range (Fig.~\ref{Hc2}a).
Data from two other cuprates are added to this plot.
First
YBa$_2$Cu$_4$O$_8$
(Y124), a stoichiometric underdoped cuprate with \Tc~$= 80$~K and \Hc~$= 44$~T,
determined both directly from $\kappa(H)$ and from $\rho(H)$ via \Hvs$(T)$ \cite{Grissonnanche2014}.
Secondly Tl$_2$Ba$_2$CuO$_6$~(Tl2201), in the strongly overdoped region.
Defining $p$ from \Tc~in all three materials using the same conversion curve
(with \Tc$^{\rm max} = 94$~K), the \Hc~data are seen to fall on a single smooth curve of \Hc~vs $p$ (Fig.~\ref{Hc2}a).

\begin{figure}[t]
\includegraphics[width=5in, trim={0 7cm 0 7cm},clip]{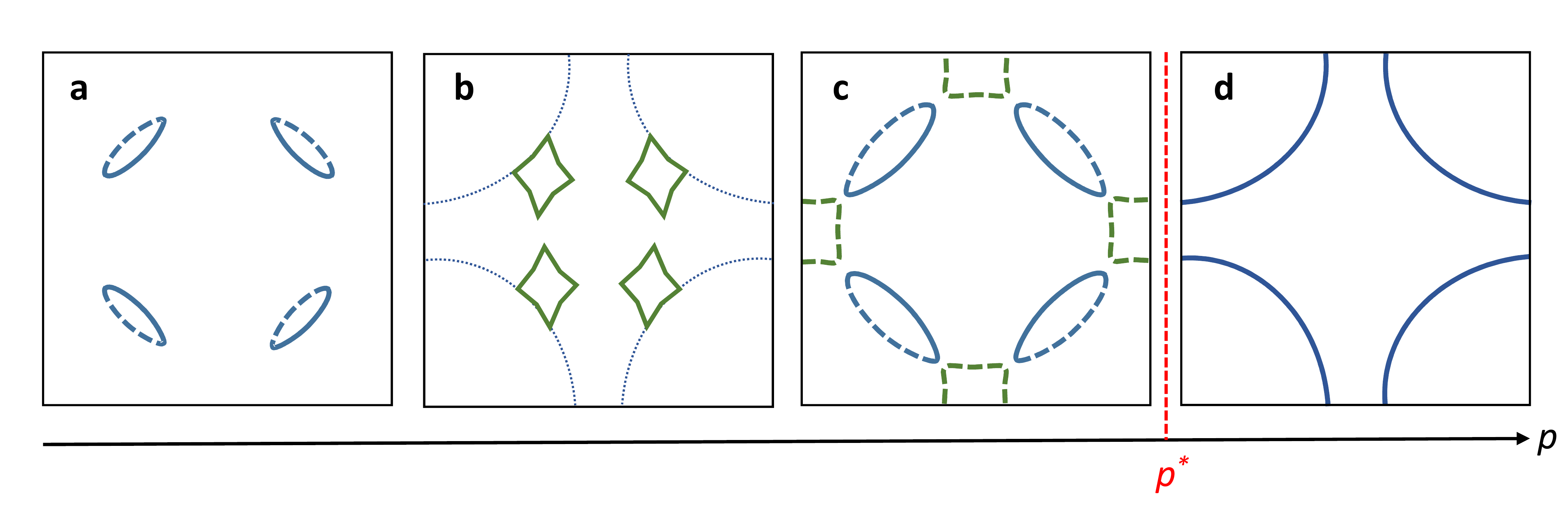}
\caption{
{\bf Sketch of Fermi surface evolution as a function of doping.}
%
At $p>$~\pstar, the Fermi surface of hole-doped cuprates is a large cylinder, either hole-like (if $p <$~\pFS), as drawn here (d),
or electron-like (if $p >$~\pFS).
At $p <$~\pstar, the topology of the Fermi surface is still unclear.
In the AF phase at low $p$ (Fig.~\ref{PhaseDiagram}), one expects small nodal hole pockets (a), containing a carrier density
$n = p$, consistent with the Hall number \nH~$\simeq p$ in YBCO and LSCO at low $p$.
If the AF phase extended up to \pstar, its Fermi surface just below \pstar~would be as sketched in (c),
with additional anti-nodal electron pockets (green).
In the CDW phase (b), the Fermi surface contains a small electron pocket, whose $k$-space location is still not established.
In one scenario \cite{HarrisonSebastian2012}, the electron pocket is located at the nodes.
}
\label{FS}
\end{figure}

\begin{marginnote}[]

\entry{LSCO}{La$_{2-x}$Sr$_x$CuO$_4$}
\entry{Nd-LSCO}{La$_{1.6-x}$Nd$_{0.4}$Sr$_x$CuO$_4$}
\entry{Eu-LSCO}{La$_{1.8-x}$Eu$_{0.2}$Sr$_x$CuO$_4$}
\entry{Hg1201}{HgBa$_2$CuO$_{6+\delta}$}

\end{marginnote}

The $H-p$ diagram of Fig.~\ref{Hc2}a is our road map:
it tells us how strong a field is required to remove superconductivity (in those three materials)
%
and it is a fingerprint of the underlying ground state.
With decreasing $p$ from the right, the striking two-peak structure in \Hc$(p)$ is shaped by the following sequence of phases:
it first rises in the strange metal phase (sec.~6),
until its highest point at \pstar, the onset of the pseudogap phase (sec.~5),
below which it drops down to a local minimum where CDW order is strongest (sec.~4),
and rises again as CDW weakens, to reach a second peak where CDW order gives way to
incommensurate SDW order, then
dropping to zero where the phase of commensurate AF order sets in (Fig.~\ref{PhaseDiagram}a).


Given that the maximal field achievable today in pulsed magnets is 100 T,
this means that superconductivity cannot currently be suppressed down to $T = 0$ in pure YBCO in the range $0.155 < p < 0.21$.
As shown earlier by Zn substitution \cite{Tallon1997}, superconductivity is most robust right around the pseudogap critical point, \pstar~$= 0.19$.
Nevertheless, applying 80 T enables one to study the normal state in that range down to at least 40~K \cite{Badoux2016}.

In LSCO,
where  \Tc$^{\rm max} \simeq 40$~K (vs 94~K in YBCO),
 \Hc~is roughly 2 times smaller than in YBCO, with a maximal value of 60-70~T.
The peak position is at $p \simeq 0.17$, close to \pstar~$\simeq 0.18-0.19$.
In Nd-LSCO, \Hc~is lower still, with a maximal
value of $\simeq 15$~T,
for a maximal \Tc~of 20~K.
The peak position is now at $p \simeq 0.22$, close to \pstar~$\simeq 0.23$.
Because of their lower effective mass \mstar,
electron-doped cuprates have a much lower \Hc~$\propto$~(\mstar)$^2$,
with a peak value of \Hc~$\simeq 10$~T
for a maximal \Tc~$\simeq 20$~K \cite{Tafti2014}.
Unless otherwise stated, the data shown in the figures and discussed in the text
are taken in magnetic fields sufficiently high to suppress superconductivity.

\section{FERMI LIQUID AT HIGH DOPING}


\begin{marginnote}[]
\entry{FL}{Fermi liquid}
\entry{WF}{Wiedemann-Franz law}
\entry{\pFS}{Lifshitz transition, where the Fermi surface changes topology from hole-like to electron-like}
\end{marginnote}

\begin{figure}[t]
\includegraphics[width=5in]{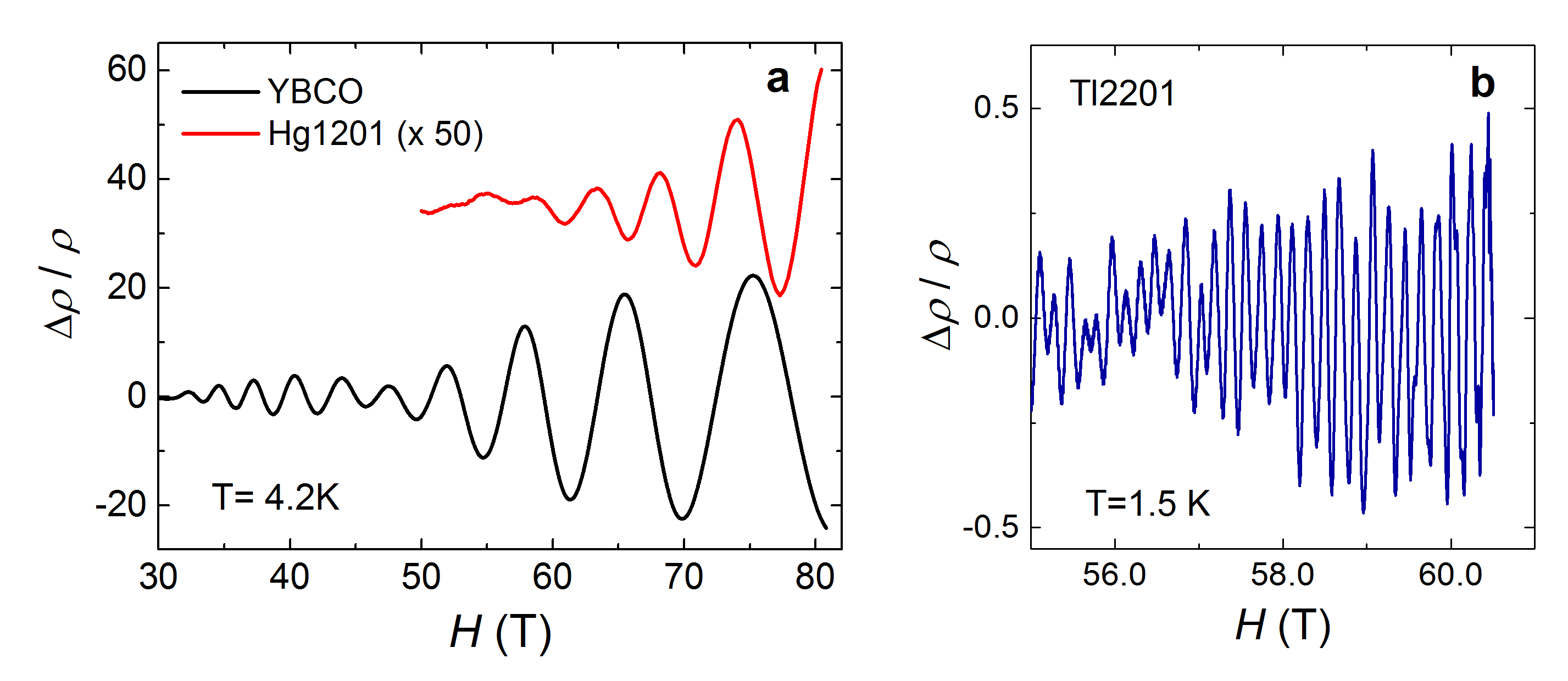}
\caption{
{\bf Quantum oscillations in cuprates.}
{\bf a)}
Underdoped YBCO ($p = 0.11$, \Tc~$=62$~K)~(black \cite{Vignolle2013})
and
Hg1201 ($p \simeq 0.1$, \Tc~$=72$~K) (red \cite{Barisic2013}, $\times 50$).
{\bf b)}
Overdoped Tl2201 ($p \simeq 0.3$,
\Tc~$\approx$~10~K)~\cite{Vignolle2008}.
}
\label{QOs}
\end{figure}

The properties of strongly overdoped cuprates are those of a Fermi liquid:
they obey the Wiedemann-Franz (WF) law \cite{Proust2002} and, beyond the superconducting dome (Fig.~\ref{PhaseDiagram}a),
their resistivity goes as $\rho \sim T^2$ in the limit $T \to 0$, as seen in Tl2201 \cite{Manako1992} and LSCO \cite{Nakamae2003}.
%
%
%
%
%

The Fermi surface of overdoped Tl2201 has been fully characterized.
It is a single large hole-like cylinder (Fig.~\ref{FS}d),
as first determined experimentally by ADMR \cite{Hussey2003} and ARPES \cite{Plate2005},
and then by quantum oscillations detected in \emph{c}-axis resistance (Fig.~\ref{QOs}b) and torque measurements
\cite{Vignolle2008}.
%
%
The oscillation frequency, $F = 18100$~T, converts to a FS area $A_F$
in excellent agreement with the $k$-space area deduced from ADMR
and ARPES.
In 2D, the Luttinger sum rule requires that the carrier density $n = 2 A_F / (2\pi)^2 = F / \phi_0$, where  $\phi_0 = h/2e$ is the flux quantum.
The  frequency measured in Tl2201 corresponds to a carrier density (per Cu atom) of $n$ = 1.3 = 1 + $p$,
%
in good agreement with the Hall number \nH~$\simeq 1.3$ obtained from Hall effect measurements at low temperature \cite{Mackenzie1996},
and with band structure calculations.
Note that at high $p$, hole-doped cuprates eventually undergo a Lifshitz transition, at \pFS, where their FS becomes electron-like.
In LSCO and Nd-LSCO, \pFS~$\simeq 0.18$ and 0.23, respectively.
%

QO measurements in
Tl2201 at $p \simeq 0.3$ yield an effective mass \mstar~$= 5.2 \pm 0.4~m_0$ \cite{Bangura2010},
where $m_0$ is the free electron mass.
%
Band structure calculations for Tl2201 obtain a bare band mass of 1.2 $m_0$ \cite{Singh1992},
implying that electron-electron interactions are
significant
when superconductivity first emerges (Fig.~\ref{PhaseDiagram}a).
For a quasi-2D FS, \mstar~is directly related to the electronic specific heat coefficient $\gamma$, via $\gamma = (\pi N_{\rm A} k_B^2 a^2 / 3\hbar^2)$~\mstar,
where $k_{\rm B}$ is the Boltzmann constant, $N_{\rm A}$ is Avogadro's number, and $a$ is the in-plane lattice constant.
In Tl2201 at $p \simeq 0.3$,
\mstar~deduced from QOs yields $\gamma = 7.6 \pm 0.6$~mJ/K$^2$ mol,
in good agreement with $\gamma = 6.6 \pm 1$~mJ/K$^2$ mol measured directly in polycrystalline Tl2201 \cite{Bangura2010}.
%
%
%
In LSCO at $p = 0.33$, $\gamma = 6.9 \pm 1$~mJ/K$^2$ mol \cite{Nakamae2003}.
The fact that two very different cuprates, Tl2201 and LSCO, have the same value of $\gamma$ at $p \simeq 0.3$ strongly suggests that
this value is generic to hole-doped cuprates in the FL phase at high doping.


\section{CHARGE-DENSITY-WAVE PHASE}

\begin{figure}[t]
\includegraphics[width=4.5in]{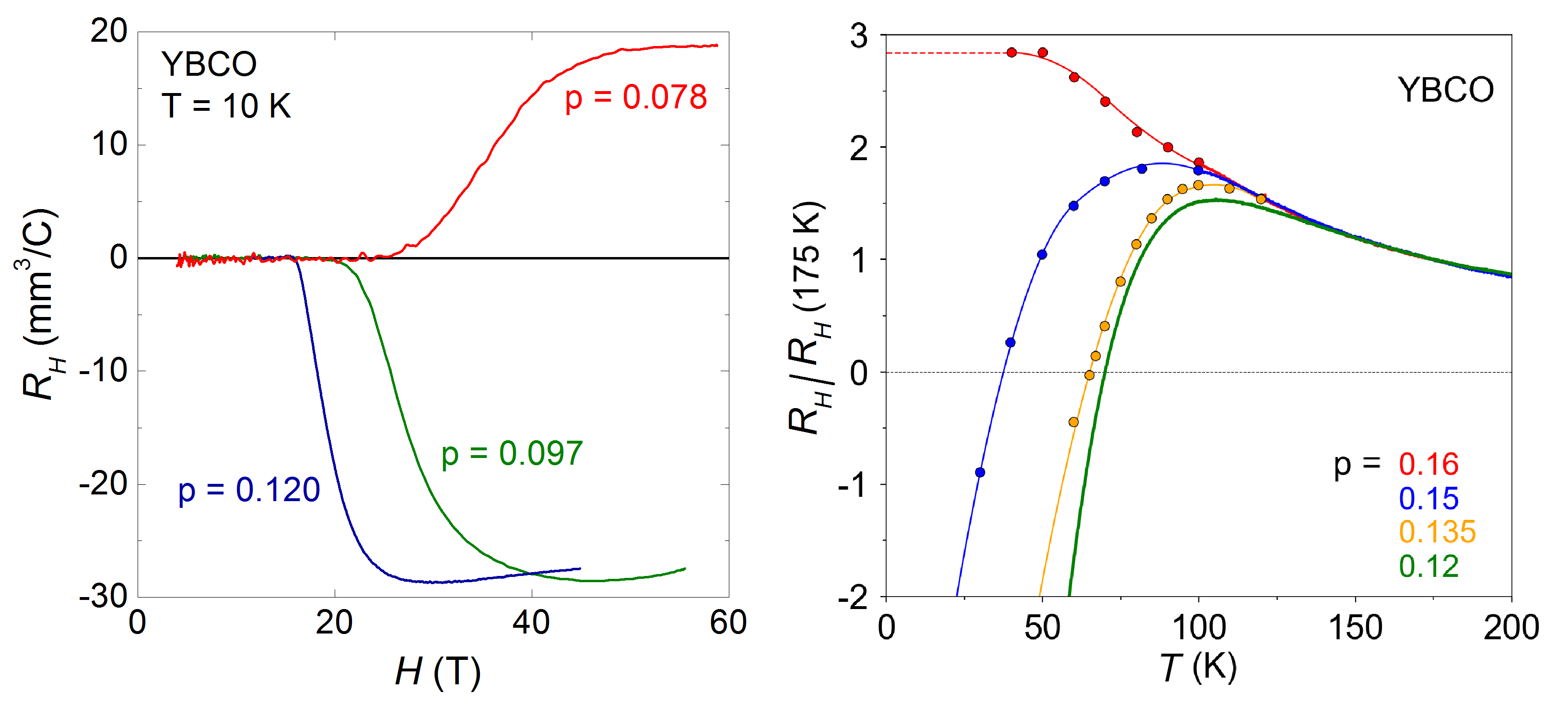}
\caption{
{\bf Fermi-surface reconstruction by CDW order as seen in the Hall coefficient.}
{\bf a)}
Field dependence of \RH~in YBCO, at three dopings as indicated.
In the normal state at high $H$, \RH~is negative at low $T$ when CDW order is present, at $p > 0.08$,
%
evidence for a small electron pocket in the FS when it is reconstructed by CDW order.
From ref.~\cite{LeBoeuf2011}.
{\bf b)}
Temperature dependence of the normal-state \RH~in YBCO, measured in high fields,
at four dopings as indicated.
\RH~at $T \to 0$ changes suddenly from negative at $p = 0.15$ to positive at $p = 0.16$,
showing that the electron pocket disappears abruptly between those two dopings.
The interval where CDW order exists in YBCO at $T=0$ is therefore $0.08 < p < 0.16$.
At $p=0.16$, the large positive \RH~is a signature of the low carrier density in the pseudogap phase,
below \pstar~$\simeq 0.19$
(Fig.~\ref{QCP}b).
From ref.~\cite{Badoux2016}.
}
\label{Hall}
\end{figure}

Since the discovery of 'stripe order' -- intertwined charge and spin modulations -- in Nd-LSCO
by neutron diffraction \cite{Tranquada1995},
followed by the discovery of charge modulations by STM in
%
Bi2212 \cite{Hoffman2002}, Ca$_2$CuO$_2$Cl$_2$  \cite{Hanaguri2004} and Bi2201 \cite{Wise2008},
it has been found that charge order
is a generic property of underdoped cuprates.
The first clue for some density wave order in YBCO came from quantum oscillations at $p \simeq 0.1$ \cite{Doiron-Leyraud2007} (Fig.~\ref{QOs}a),
in combination with a negative Hall coefficient at low temperature \cite{LeBoeuf2007} (Fig.~\ref{Hall}a).
%
%
First detected in the resistance, QOs in YBCO have since been observed in the magnetization \cite{Jaudet2008}, Nernst and Seebeck coefficients \cite{Laliberte2011,Doiron-Leyraud2015}, specific heat \cite{Riggs2011}, and thermal conductivity \cite{Grissonnanche2014}.
The main frequency $F \approx 530$~T (Fig.~\ref{QOs}a) corresponds to an extremal area $A_F$ only 2~$\%$ of the first Brillouin zone,
while the large orbit in overdoped Tl2201 represents 65~$\%$ (Fig.~\ref{QOs}b).
Fig.~\ref{Hall}b shows the normal-state Hall coefficient in YBCO.
At low $T$,
\RH~is deeply negative in the doping interval $0.08 < p < 0.16$
\cite{LeBoeuf2011,LeBoeuf2007},
precisely where QOs have been observed \cite{SebastianPNAS2010,Ramshaw2015}.
The combination of QOs and a negative \RH~is a strong indication for the presence of a small closed electron pocket in the Fermi surface, also
consistent with the magnitude and negative sign of the Seebeck coefficient \cite{Laliberte2011,Chang2010}.
The most natural interpretation for the presence of a small electron pocket is a FS reconstruction induced by some density wave that breaks translational symmetry.
High-field NMR experiments in YBCO
later
showed that this broken symmetry comes from a long-range CDW order \cite{Wu2011},
in the absence of SDW modulations --
thereby
distinct from
the previously known 'stripe order'.

\begin{marginnote}[]
\entry{Bi2212}{Bi$_2$Sr$_2$CaCu$_2$O$_{8 + \delta}$}
\entry{Bi2201}{Bi$_2$Sr$_2$CuO$_{6 - \delta}$}

\end{marginnote}

Two distinct charge orders are detected in YBCO by X-ray diffraction.
First, a 2D short-range (but static) bidirectional CDW appears well above \Tc~in the doping range $0.08 < p < 0.16$ \cite{Blanco-Canosa2014,Hucker2014}.
Charge modulations are incommensurate with an in-plane correlation length of at most 20 lattice constants.
The second CDW, originally detected using high field NMR, appears below \Tc~and above a threshold field that is doping dependent,
as shown by high-field sound velocity measurements
\cite{Laliberte2018}.
Recent X-ray measurements in high field \cite{Gerber2015,Chang2016} have shown that it is a 3D ordered state with in-plane CDW modulations along the $b$ direction only (but with the same period as the 2D CDW). Compared to the 2D short range CDW, the in-plane and \emph{c}-axis correlation lengths are greatly enhanced and the former extends to $\simeq 60$ lattice constants.
Both CDWs coexist at low temperature in the exact same doping range \cite{Laliberte2018}.
%

Although with a much shorter correlation length than in YBCO,
bidirectional CDW
order
has also been detected by X-ray diffraction in several other cuprates,
namely Hg1201 \cite{Tabis2014}, LSCO \cite{Croft2014,Thampy2014},
Bi2212 \cite{DaSilvaNeto2014} and Bi2201 \cite{Comin2014} -- demonstrating that CDW order is a universal tendency of hole-doped cuprates.
And it has an unusual $d$-wave form factor \cite{STM-dDW}.
Note that low-frequency quantum oscillations ($F \approx 840$~T) \cite{Barisic2013} (Fig.~\ref{QOs}a)
and negative Hall and Seebeck coefficients \cite{Doiron-Leyraud2013} are also observed in Hg1201
despite the much shorter correlation length and the lack (so far) of a field-induced 3D CDW.
%

%
The exact mechanism of FS reconstruction by the CDW is still debated.
A biaxial charge order with wavevectors ($Q_x$, 0) and (0, $Q_y$) will create a small electron-like pocket located at the nodes \cite{HarrisonSebastian2012} (Fig.~\ref{FS}b).
A unidirectional CDW can create an electron-like pocket in the presence of nematic order \cite{Yao2011}, but it would be located at the antinodes.
%
The $k$-space location of the electron pocket is unknown, and it is unclear at the moment if there are additional sheets in the reconstructed FS.
In YBCO at $p = 0.11-0.12$, the QO mass \mstar~accounts for only 2/3 of the measured specific heat $\gamma$ \cite{Klein2018,Marcenat2016}.
Whether the missing 1/3 is due to a CuO chain band or some other (open or closed) sheet associated with the CuO$_2$ planes remains to be seen.
%
The presence of another sheet could account for the anomalous doping dependence of the Seebeck coefficient in YBCO \cite{Laliberte2011,Doiron-Leyraud2015}
and for the anomalous magnitude of the $T^2$ resistivity in Y124 \cite{Proust2016}.
%

The relation between CDW order and the pseudogap phase is the subject of ongoing research,
but one thing is now clear:
at $T \to 0$, the CDW phase ends at a critical doping \pcdw~which is distinctly lower than \pstar.
This has been shown for YBCO \cite{Badoux2016} (see Fig.~\ref{Hall}b) and LSCO \cite{BadouxPRX2016},
and it is also clear for Eu-LSCO \cite{Laliberte2011} and Nd-LSCO \cite{Collignon2017,Daou2009} where
\RH~and $S$
at $T \to 0$
are no longer negative at $p = 0.20$ and above, while \pstar~$=0.23$ (Fig.~\ref{PhaseDiagram}b).

%


A remarkable aspect of the CDW phase is that its ground state is a Fermi liquid,
even though it lives completely inside the pseudogap phase (Fig.~\ref{PhaseDiagram}a) -- a phase with seemingly incoherent quasiparticles and
mysterious disconnected Fermi arcs.
Indeed, in the CDW phase of YBCO, not only are QOs observed that obey the standard Lifshitz-Kosevich formula \cite{Sebastian2010} --
proof of a closed FS and coherent quasiparticles --
but
the Wiedemann-Franz law is obeyed \cite{Grissonnanche2016} and
the in-plane resistivity has the classic $T^2$ dependence at $T \to 0$ \cite{LeBoeuf2011,Proust2016}.


%
%

\section{PSEUDOGAP PHASE}


\begin{marginnote}[]
\entry{DOS}{Density of states ($N_{\rm F}$)}
\entry{$\delta E$}{Condensation energy}
\entry{$H_{\rm c1}$}{Lower critical field}
\entry{$\gamma$}{Residual linear term in the specific heat, $C(T)$ at $T=0$, purely electronic}
\end{marginnote}

\begin{figure}[t]
\includegraphics[width=5in]{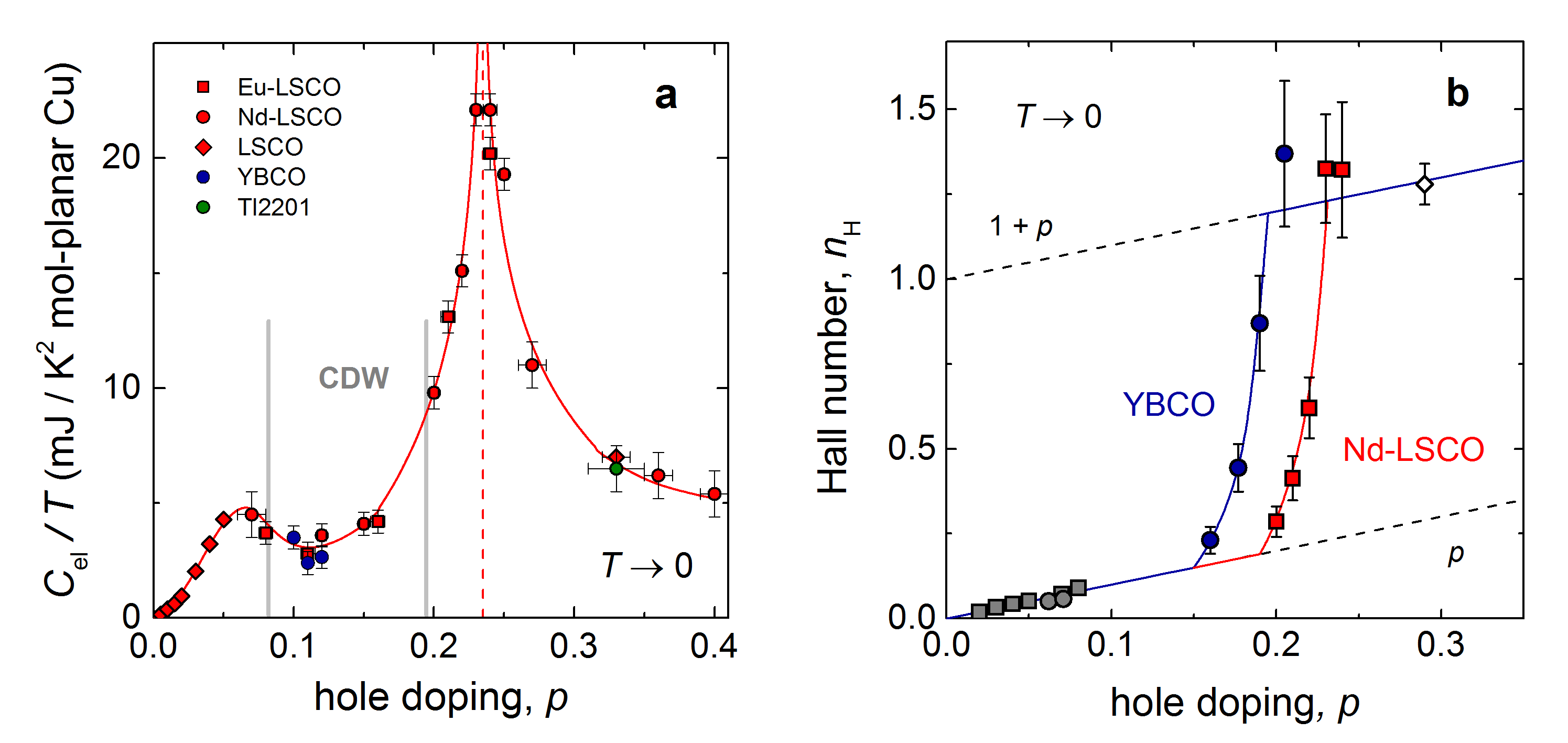}
\caption{
{\bf Across the quantum critical point.}
{\bf a)}
Normal-state electronic specific heat in the $T=0$ limit as a function of doping,
plotted as \CelT~vs $p$ (red symbols)
in Eu-LSCO (squares), Nd-LSCO (circles) and LSCO (diamonds).
From ref.~\cite{Michon2018}.
We also show \CelT~in YBCO (blue dots \cite{Klein2018}) and in Tl2201 (green dot \cite{Wade1994}).
The vertical grey lines mark the limits of the CDW phase in Nd-LSCO,
between $p = 0.08$ and $p \simeq 0.19$.
{\bf b)}
Normal-state Hall number \nH~($=V / e$~\RH) in the $T=0$ limit as a function of doping,
in YBCO (blue circles \cite{Badoux2016}, \pstar = 0.19) and Nd-LSCO (red squares \cite{Collignon2017}, \pstar = 0.23).
We also show \nH~in LSCO (grey squares \cite{Ando2004}) and YBCO (grey circles \cite{Segawa2004}) at low doping, and \nH~in Tl2201 (white diamond \cite{Mackenzie1996}) at high doping.
}
\label{QCP}
\end{figure}

The two traditional signatures of the pseudogap phase are:
1) a loss of density of states (DOS) below \pstar;
2) the opening of a partial spectral gap below \Tstar,
seen by ARPES
(Figs.~\ref{PhaseDiagram}c,~\ref{PhaseDiagram}d)
and optical conductivity, for example.
%
Here we summarize recent high-field measurements of the specific heat in the LSCO family
\cite{Michon2018}
showing that there is a large mass enhancement at \pstar.
The new data show that the pseudogap does not simply cause a loss of DOS below \pstar;
instead, there is huge peak
in the DOS at \pstar~(Fig.~\ref{QCP}a) -- much larger than expected from a van Hove singularity
\cite{Michon2018,Horio2018}.
%
%
We then show how high-field measurements of the Hall coefficient
reveal a new signature of the pseudogap phase -- a rapid drop in the carrier density,
at \pstar
(Fig.~\ref{QCP}c).
%
These new properties alter profoundly our view of the pseudogap phase,
and of the strange metal just above it (sec.~6).
%

\subsection{Density of states}

\subsubsection{Condensation energy}

One way to access the DOS, $N_{\rm F}$, is via the superconducting condensation energy $\delta E$,
since $\delta E = N_{\rm F} \Delta_0^2 / 4$, where $\Delta_0$ is the $d$-wave gap maximum.
Experimentally,
and in the framework of BCS theory, $\delta E$ can be measured using the upper and lower critical fields, \Hc~and $H_{\rm c1}$, to
get the thermodynamic field $H_{\rm c}$
via $H_{\rm c}^2 = H_{\rm c1} H_{\rm c2} / ({\rm ln}(\kappa) + 0.5)$, given that $\delta E = H_{\rm c}^2 / 2 \mu_0$.
In Fig.~\ref{Hc2}b, we plot $\delta E /$\Tc$^2$ vs $p$ thus obtained for YBCO~\cite{Grissonnanche2014}.
We see that $\delta E /$\Tc$^2 \propto N_{\rm F}$ drops by a factor 8-9 between $p = 0.18$ and $p = 0.1$,
in  agreement with the drop reported earlier from an analysis of specific heat data
measured in low fields up to $T >$~\Tc~in
YBCO \cite{Luo2000} and Bi2212 \cite{Loram2000}.
Note that $\sim 2/3$ of this drop has taken place {\em before} the onset of the CDW phase
(grey vertical lines in Fig.~\ref{Hc2}), so that it is indeed a property of the 'pure' pseudogap phase.
CDW order causes an additional depletion of DOS near $p \simeq 0.12$
(Fig.~\ref{Hc2}b).

A more direct way to access $N_{\rm F}$ is to measure the normal-state specific heat coefficient
$\gamma$ in the $T=0$ limit,
given by $\gamma = (2\pi^2 k_{\rm B}^2/3)~N_{\rm F}$
in the standard theory of metals.
This requires a field large enough to fully suppress superconductivity,
which has recently been achieved for YBCO at $p = 0.10 - 0.12$ \cite{Klein2018}
and for Nd-LSCO across the phase diagram \cite{Michon2018} (sec.~5.1.2).
In YBCO at $p = 0.10$ (\Tc~$=56$~K), the $H$-dependent part of $\gamma$ is measured to be
$\delta \gamma = \gamma(H_{\rm c2}) - \gamma(0) = 3.75$~mJ/K$^2$mol-Cu \cite{Klein2018}.
We can use this to estimate $\delta E /$\Tc$^2$ via the relation
$\delta E /$\Tc$^2 = (3/8\pi^2)(\Delta_0/k_{\rm B}T_{\rm c})^2 \delta \gamma$,
taking
$\Delta_0/k_{\rm B}T_{\rm c} = 2.8$,
consistent with various experimental estimates
\cite{Hufner2005,Peets2007}.
The result is plotted as a blue dot in Fig.~\ref{Hc2}b -- in
good
 agreement with the value
derived from \Hc~
and $H_{\rm c1}$.

%

Having anchored the magnitude of $\delta E /$\Tc$^2$ in YBCO, at $p=0.1$,
the observed rise by a factor 8-9 up to $p=0.18$ (Fig.~\ref{Hc2}b) therefore implies that  $N_{\rm F}$ becomes very large
at \pstar~$\simeq 0.19$, namely $\delta \gamma \simeq 30$~mJ/K$^2$mol-Cu, implying that \mstar$ \simeq 20~m_0$.
The beginning of this rise was detected as an increase in \mstar~measured via QOs in YBCO \cite{Ramshaw2015}.
The question is: what happens above \pstar?

Measurements on Tl2201 and LSCO show that the DOS must undergo a major drop above \pstar.
%
Indeed, in overdoped Tl2201 with \Tc~$\simeq 20$~K,
\mstar~$=5.2~m_0$~from QOs (sec. 3) -- a factor 4 down from $\simeq 20~m_0$ in YBCO at \pstar.
%
%
%
This drop in $N_{\rm F}$ {\em above} \pstar~is
an important
feature of cuprate behavior that has gone unnoticed until recently.


\begin{figure}[t]
\includegraphics[width=6.2in]{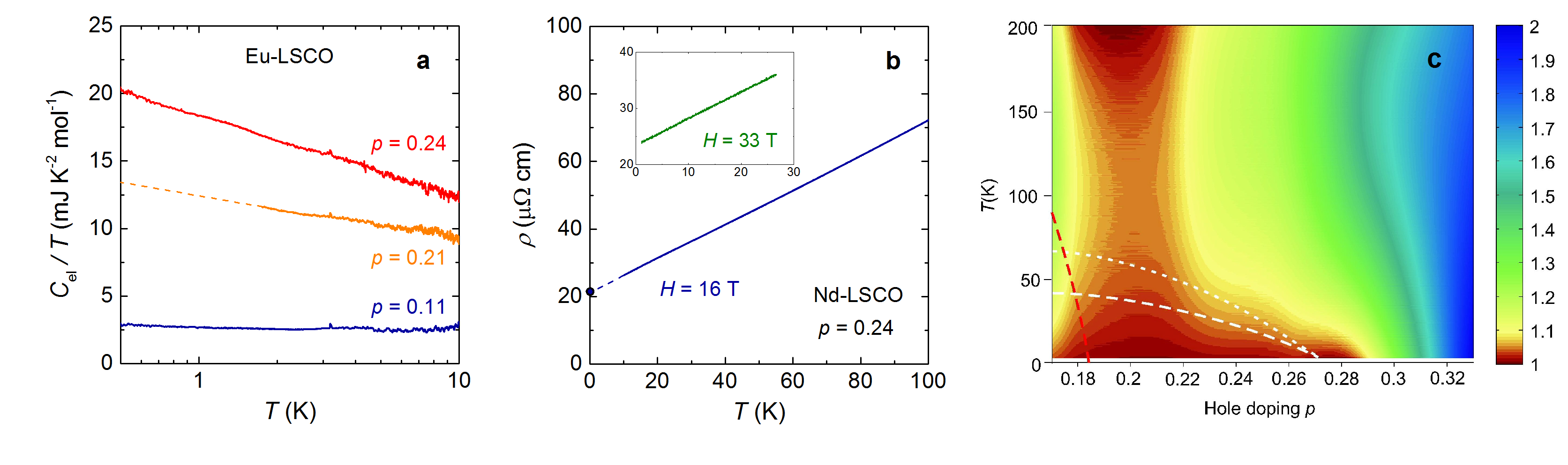}
\caption{
{\bf At the quantum critical point.}
{\bf a)}
Temperature dependence of the normal-state electronic specific heat, plotted as $C_{\rm el} / T$
vs log$(T)$,
in Eu-LSCO at three dopings.
At $p = 0.24 \simeq$~\pstar~(red),
$C_{\rm el} / T \sim$~log$(1/T)$
down to the lowest temperature, the thermodynamic signature of a QCP.
From ref.~\cite{Michon2018}.
{\bf b)}
Temperature dependence of the normal-state resistivity
in Nd-LSCO at $p = 0.24$, at $H=16$~T (main panel; \cite{Michon2018b}) and $H = 33$~T (inset; \cite{Daou2009}).
At $p = 0.24 \simeq$~\pstar,
$\rho \sim T$
down to the lowest temperature, the transport signature of a QCP.
The blue dot on the $y$ axis is the thermal resistivity of the same sample extrapolated to $T=0$, $1 / \kappa_0$,
from data measured down to 0.1~K at $H=15$~T, expressed as $L_0 T / \kappa_0$ (sec.~5.3.4).
{\bf c)}
Temperature-doping map of the temperature exponent $n$ in the resistivity $\rho \sim T^n$ of LSCO.
From ref.~\cite{Cooper2009}.
}
\label{NFL}
\end{figure}

\subsubsection{Specific heat}


The presence of a peak in the DOS of cuprates at \pstar~can be established by measuring
the normal-state specific heat in the $T \to 0$ limit across the phase diagram in one and the same material.
This has recently been done on Nd-LSCO \cite{Michon2018}, a single-layer cuprate with a low \Tc~($\leq 20$~K) and a low \Hc~($\leq 15$~T),
so that 15 T is sufficient to suppress superconductivity at any doping (in bulk thermal measurements).
The phase diagram of Nd-LSCO is shown in Fig.~\ref{PhaseDiagram}b, with
the boundary \Tstar$(p)$ of its pseudogap phase delineated by both transport \cite{Collignon2017,Daou2009}
and ARPES \cite{Matt2015} measurements.
The critical doping at which the pseudogap phase ends is \pstar~$= 0.23 \pm 0.01$.

In Fig.~\ref{QCP}a, the normal-state electronic specific heat \Cel~of Nd-LSCO is plotted  as \CelT~vs $p$,
at $T = 0.5$~K (red circles, \cite{Michon2018}), along with corresponding data taken on Eu-LSCO (squares, \cite{Michon2018})
and LSCO (diamonds, \cite{Komiya2009,Nakamae2003}).
The agreement amongst the three cuprates is excellent.
The red curve gives us the detailed evolution of \CelT~in the $T=0$ limit across the entire phase diagram,
from the Mott insulator at $p=0$ all the way to the Fermi liquid up to $p = 0.4$.
%
The dominant feature is a huge peak at \pstar.

This establishes, by raw data from a measurement that directly gives the DOS, taken on three materials (LSCO, Eu-LSCO and Nd-LSCO),
the presence of a large peak in the DOS of cuprates that we
inferred from our piecemeal construction
based on two other materials (YBCO and Tl2201).
The comparison can be anchored quantitatively by adding the specific heat data for YBCO (blue dots) and Tl2201 (green dot)
in Fig.~\ref{QCP}a. We see that the values of $\gamma =$~\CelT
per CuO$_2$ plane
are similar in YBCO, Nd-LSCO and Eu-LSCO at
$p \simeq 0.12$ and again in LSCO, Nd-LSCO and Tl2201 at $p \simeq 0.3$, showing that the DOS is essentially
the same in all cuprates to the left and right of \pstar.
Importantly, $\gamma$ has the same magnitude at $p \simeq 0.07$ and at $p = 0.4$,
so that the pseudogap
phase is {\em not} characterized by a loss of DOS relative to the overdoped FL,
but by a peak in the DOS at \pstar.
%


Early specific heat measurements on polycrystalline samples of LSCO in which Zn impurities were added
to suppress superconductivity had already revealed a peak in $\gamma$ vs $p$,
centered at $p \simeq 0.2$ \cite{Momono1994}, a broadened version of the peak seen in pure samples of Nd-LSCO (Fig.~\ref{QCP}a).
This peak
was
attributed to the van Hove singularity in the band structure of LSCO, at \pFS~$\simeq 0.2$,
but calculations
now
contradict
this interpretation (for both LSCO \cite{Horio2018} and Nd-LSCO \cite{Michon2018}),
 as the heavy disorder
and strong 3D dispersion completely flatten the singularity in the DOS.
%
%
%
%

\subsection{Quantum critical point}


The sharp peak in \CelT~vs $p$ (Fig.~\ref{QCP}a) is a classic thermodynamic signature of a QCP.
Indeed, a very similar peak has been observed in the iron-based superconductor BaFe$_2$(As$_{1-x}$P$_{x}$)$_2$,
at the QCP where its AF order vanishes vs $x$ \cite{Walmsley2013}.
A second thermodynamic signature of a QCP is a logarithmic divergence of \CelT~as $T \to 0$,
as observed in the heavy-fermion metal CeCu$_{6-x}$Au$_x$
at the QCP where its AF order vanishes vs $x$ \cite{Lohneysen1994}.
As seen in Fig.~\ref{NFL}a, this is indeed the behavior measured in Eu-LSCO
at \pstar,
and also in Nd-LSCO~\cite{Michon2018}.
The third classic signature of a QCP is a $T$-linear resistivity as $T \to 0$,
found in both BaFe$_2$(As$_{1-x}$P$_{x}$)$_2$ and CeCu$_{6-x}$Au$_x$ at their  QCP,
and indeed in Nd-LSCO at \pstar~(Fig.~\ref{NFL}b).
The striking $T$-linear resistivity of overdoped cuprates is discussed in sec.~6.

All this is compelling empirical evidence that the pseudogap critical point is a QCP.
But what is missing so far is the detection of a diverging length scale associated with quantum criticality in hole-doped cuprates.
In sec.~5.4, we examine the obvious question:
What order ends at that quantum phase transition, if any?

\subsection{Carrier density}


Having established how the DOS in the ground state of cuprates evolves
from Fermi liquid at $p > 0.3$ to Mott insulator at $p=0$,
we now look at a separate yet equally important property of a metal,
its carrier density, $n$.

\subsubsection{Hall coefficient}

The carrier density can be accessed by measuring the Hall coefficient \RH.
%
In the following, we assume that the Hall number \nH~gives the carrier density $n$ in the $T$ = 0 limit.
This is the case in
Tl2201 at $p \simeq 0.3$, \RH~in the $T=0$ limit is such that \nH~$\simeq 1 + p$,
consistent with $n = 1 + p$ measured by QOs, ADMR and ARPES (Sec.~3).
%
%
At the other end of the phase diagram, Hall measurements in YBCO \cite{Segawa2004} and LSCO \cite{Ando2004} give
\nH~$\simeq p$ at low $T$ up to $p \simeq 0.08$
(Fig.~\ref{QCP}c).
%
A carrier density $n = p$ is what is required by the Luttinger rule in the presence of
commensurate AF order with a wavevector $(\pi, \pi)$, since the AF
Brillouin zone contains one fewer electron (Fig.~\ref{FS}a).
Since commensurate AF order prevails in YBCO up to $p = 0.05$ (Fig.~\ref{PhaseDiagram}a)
and in LSCO up to $p \simeq 0.02$, followed by incommensurate SDW order up to $p \simeq 0.08$
and $p \simeq 0.12$, respectively, the fact that \nH~$\simeq p$ up to $p \simeq 0.08$
is perhaps understandable.
The question is :
how exactly does the ground state evolve, with decreasing $p$, from a metal with $n = 1 + p$ to another metal with $n = p$?

This question was answered by two recent high-field measurements of \RH, summarized in Fig.~\ref{QCP}b.
In both YBCO \cite{Badoux2016} and Nd-LSCO \cite{Collignon2017} with decreasing $p$,
a rapid drop in \nH~starts in tandem with the opening of the pseudogap (\pstar~$\simeq$~0.19 in YBCO and \pstar~$\simeq$~0.23 in Nd-LSCO).
The drop in \nH~is equally rapid and deep in the two very different materials, showing that it must reflect a generic underlying property
of hole-doped cuprates.
%
%
(Note that earlier high-field studies of \RH~in Bi2201 \cite{Balakirev2003} and LSCO \cite{Balakirev2009}
do not show as clean a drop in \nH.
We attribute this to
the contaminating effect of CDW order,
which produces a drop in \RH, and therefore an apparent increase in \nH, at low $T$.
%
Also, in Bi2201, the Hall study does not extend up to \pstar.)
%

The simplest interpretation of the Hall data in YBCO and Nd-LSCO is a transition at \pstar~that causes the carrier density to go
from $n = 1+p$ above \pstar~to $n = p$ below,
consistent with AF order with $Q = (\pi, \pi)$,
for example \cite{Storey2016}.
%
However, it is in principle possible that \RH~could increase because of a change in FS curvature
rather than a change in FS volume,
as in the case of a nematic transition \cite{Kivelson}.
To confirm that carriers are indeed lost below \pstar, we turn to three other transport properties.

\subsubsection{Thermopower}

In the $T=0$ limit, the Seebeck coefficient $S$ of a metal is equal to the entropy per carrier,
given by the simple relation $S/T \sim \gamma / n$,
validated in many different families of materials \cite{Behnia2004}.
The Seebeck coefficient of Nd-LSCO was measured at low $T$ above and below \pstar~\cite{DaouPRB2009}.
In going from $p = 0.24 >$~\pstar~to $p = 0.20 <$~\pstar,
$S/T$ at low $T$ ($\leq 10$~K) increases by a factor 4-5 \cite{DaouPRB2009}.
Given that $\gamma$ drops, this necessarily implies that $n$ must decrease significantly below \pstar.
We conclude that there is definitely a drop in carrier density upon entering the pseudogap phase.
%

%

\begin{figure}[t]
\includegraphics[width=6.3in]{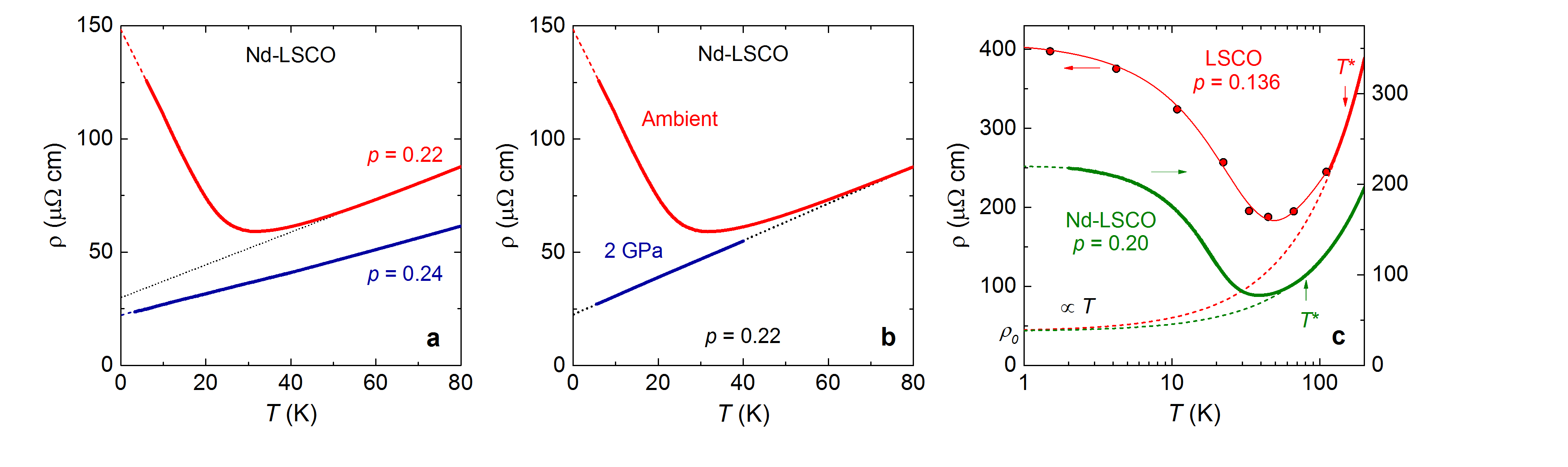}
\caption{
{\bf Resistivity upturns in the pseudogap phase.}
{\bf a)}
Temperature dependence of the normal-state resistivity in Nd-LSCO,
at $p = 0.22 <$~\pstar (red) and $p = 0.24 >$~\pstar (blue).
From ref.~\cite{Collignon2017}.
{\bf b)}
The upturn
at $p = 0.22$ can be entirely removed by applying a pressure $P = 2$~GPa,
showing that pressure lowers \pstar~below 0.22.
From ref.~\cite{Doiron-Leyraud2017}.
{\bf c)}
Normal-state resistivity of
LSCO at $p = 0.136$~\cite{Laliberte2016} (red) and
Nd-LSCO at $p = 0.20$~\cite{Collignon2017} (green),
plotted as $\rho$ vs log$(T)$.
%
Dashed lines are extensions of a linear fit to high-$T$ data.
The vertical arrows mark the pseudogap temperature \Tstar. From ref.~\cite{Laliberte2016}.
}
\label{Upturns}
\end{figure}

\subsubsection{Resistivity}

In Fig.~\ref{Upturns}a, we show what happens to the resistivity $\rho(T)$ of Nd-LSCO
upon entering the pseudogap phase.
At $p = 0.24$, just above~\pstar$=0.23$, $\rho(T)$ is perfectly linear, from  80~K or so down to $T \to 0$ (Fig.~\ref{NFL}b).
At $p = 0.22$, just below~\pstar, $\rho(T)$ exhibits a large upturn at low $T$
but saturates to a finite value in the $T$ = 0 limit.
The upward deviation from the $T$-linear behaviour at high $T$ starts at \Tstar$\simeq 50$~K (Fig.~\ref{PhaseDiagram}b).
The increase in the normal-state resistivity at $T = 0$ corresponds to a
5-fold
 drop in conductivity.

It has recently been demonstrated that a pressure of 2~GPa lowers \pstar~in Nd-LSCO below $p=0.22$ \cite{Doiron-Leyraud2017}.
As a result, 2 GPa applied to a sample with $p=0.22$ eliminates the upturn completely, and
$\rho(T)$ then becomes perfectly $T$-linear (Fig.~\ref{Upturns}b).
We see again that $\rho(0)$ at $T=0$ changes by a
factor $\simeq 5$.
%
This large drop in conductivity below \pstar~is consistent with a loss of carrier density.

Low-$T$ upturns in the resistivity of cuprates were discovered two decades ago in LSCO,
when two different approaches were used to access the normal state at $T \to 0$.
In a first approach,
Zn impurities were used to kill superconductivity in polycrystalline samples of LSCO \cite{Momono1994}.
Resistivity measurements revealed two distinct regimes of behaviour:
for $p > 0.17$, a $T$-linear dependence below 100~K, up to at least $p = 0.22$;
for $p < 0.17$, a low-$T$ upturn, that grows with decreasing $p$.
Soon after, this phenomenology was confirmed by studies on (Zn-free) LSCO single crystals in pulsed magnetic field measurements (up to 61 T) \cite{Boebinger1996,Ando1995},
with the transition from $T$-linear to upturn occurring also at $p \simeq 0.17$.
The striking similarity between the two studies reveals two important facts.
First, the field plays no other role than to remove superconductivity, so the upturn is not a high-field effect ({\it e.g.} it is not due to magneto-resistance).
Secondly, disorder does not cause the low-$T$ upturns
(just below \pstar) -- they appear below a critical doping \pstar~where the underlying ground state changes.

For 20 years, this phenomenology was dubbed 'metal-to-insulator crossover',
but it is now clear that the change of behaviour in $\rho(T)$ is the result of a metal-to-metal transition
occurring at \pstar, characterized by a large drop in carrier density \cite{Laliberte2016},
whose key signature is a loss of carrier density $n$ \cite{Daou2009, Collignon2017, Badoux2016}.
In LSCO, the data show that \pstar$\simeq 0.18-0.19$.
That the ground state is a metal and not an insulator, nor a metal with charge localization,
can be seen in two ways.
First, by plotting $\rho(T)$ vs log$T$, as done in Fig.~\ref{Upturns}c, where we see that $\rho(T)$ saturates
as $T \to 0$ in Nd-LSCO at $p=0.20$ and LSCO at $p = 0.136$, despite the
large increase in $\rho(0)$.
%
(Of course, as in any metal, localization does occur when $k_F l_0$ approaches 1.0,
as observed in LSCO when $p < 0.1$ \cite{Boebinger1996}.)

Another
way to confirm that the pseudogap is a metal at $T=0$
is to look at thermal conductivity at very low temperature, as we now discuss.

\subsubsection{Thermal conductivity}

The thermal conductivity $\kappa(T)$ of Nd-LSCO was recently measured down to $T \simeq 0.1$~K, for dopings across \pstar~\cite{Michon2018b}.
Straightforward extrapolation of the data to $T=0$ yields the purely fermionic residual linear term $\kappa_0/T$.
In a field large enough to fully suppress superconductivity, this then gives the thermal conductivity
of the underlying ground state of this cuprate, above and below \pstar.
The data is found to accurately obey the Wiedemann-Franz (WF) law.
Indeed, at $p = 0.24$, $\kappa_0/T = L_0 / \rho(0)$, where $L_0 \equiv (\pi^2/3)(k_{\rm B}/e)^2$,
within
 an error bar of only a few percent \cite{Michon2018b}.
Here, $\rho(0)$ is obtained by a linear extrapolation of $\rho(T)$ to $T=0$, as shown in Fig.~\ref{NFL}b.
%

The WF law is also obeyed when $p <$~\pstar~\cite{Michon2018b}, showing that the pseudogap ground state, truly at $T=0$, is indeed a metal.
The fermionic quasiparticles of that state are charged and they carry heat just as normal electrons do.
The
5-fold drop
 in $\kappa_0/T$ between $p = 0.24$ and $p = 0.22$ is therefore perfectly consistent with the
5-fold increase
in $\rho(0)$ (Fig.~\ref{Upturns}a), confirming it is a property of the ground state.
Moreover,
the same measurements performed in zero field, and so inside the superconducting state,
find a similarly large drop in
thermal
conductivity across \pstar, showing that the transport signatures interpreted as a drop in $n$ are not induced by the applied
magnetic field and the pseudogap phase is not affected by such fields \cite{Michon2018b}.





\subsection{Scenarios for the pseudogap phase}

To summarize, we
can list the following properties of the pseudogap phase at $T=0$,
in the absence of either superconductivity or CDW order:

\begin{enumerate}
\item Anti-nodal spectral gap opens below \pstar~(Fig.~\ref{PhaseDiagram}c)
\item Density of states decreases below \pstar~(Figs.~\ref{Hc2}b,~\ref{QCP}a)
\item Carrier density drops below \pstar, from $n \simeq 1+p$ to $n \simeq p$~(Fig.~\ref{QCP}b)
\item Wiedemann-Franz law is obeyed~(Fig.~\ref{NFL}b)
\item Transition vs $p$ (at \pstar, $T=0$), crossover vs $T$ (at \Tstar, $p <$~\pstar)
\item Signatures of quantum criticality at \pstar, with \Cel~$\sim T$log$T$ and $\rho \sim T$~(Figs.~\ref{NFL}a,~\ref{NFL}b)
\end{enumerate}


\subsubsection{AF scenario}

At the empirical level,
these signatures
are reminiscent of a
scenario of AF order in 2D for the pseudogap phase of hole-doped cuprates.
By analogy with organic conductors, heavy-fermion metals and iron-based superconductors,
but also with electron-doped cuprates.
Indeed, the phenomenology of electron-doped cuprates (see Sidebar) is
essentially the same as what we have listed here for hole-doped cuprates,
and in the former there is little doubt that a scenario of AF order and QCP is appropriate.
The reason why this is immediately reasonable in that case is this:
the AF correlation length measured by neutrons \cite{Motoyama2007} is found to increase rapidly just below the QCP
detected in transport \cite{Dagan2004} and ARPES \cite{Matsui2007}.
Conversely, the reason why an AF QCP scenario is {\em not} immediately reasonable for
hole-doped cuprates is because no long AF correlation length has been universally detected in this case.

A fundamental question remains open today:
are
AF spin fluctuations / correlations generically present in the pseudogap phase of cuprates at $T=0$
when superconductivity and CDW order are removed?

Two other empirical features suggest that AF spin fluctuations / correlations
are involved in the formation of the pseudogap phase.
The first is that $d$-wave pairing in cuprates is most likely caused by AF spin fluctuations \cite{Scalapino2012},
in both electron-doped and hole-doped, and so it is likely that those are the fluctuations
associated with the QCP located inside the \Tc~dome -- as is indeed the case for most other unconventional superconductors with a \Tc~dome.
%
The second feature is the recently proposed constraint that \pstar~$\leq p_{\rm FS}$
\cite{Doiron-Leyraud2017,Benhabib2015}.
In Nd-LSCO, the inequality \pstar~$\leq p_{\rm FS}$ was inferred from the fact that pressure lowers $p_{\rm FS}$ and \pstar~by the same amount,
as detected by the lowering of \RH~and the disappearance of the upturn in $\rho(T)$ (Fig.~\ref{Upturns}b), respectively \cite{Doiron-Leyraud2017}.
This would also explain why \pstar~is significantly lower in LSCO vs Nd-LSCO, given that $p_{\rm FS}$ also is,
in spite of their having the same \Tstar~boundary at low doping \cite{Cyr-Choiniere2018}.
The simplest intuitive way to understand why the pseudogap does not open when the Fermi surface is
electron-like is to associate the pseudogap with hot spots on the FS where it intersects the AF zone boundary.
When the Fermi surface becomes electron-like, those hot spots disappear, as no intersection is possible.
The central role of the AF zone boundary as the $k$-space organizing principle for the pseudogap and the associated transformation of the Fermi surface,
from large circle above \pstar~to small arcs below, is vividly seen in STM data on Bi2212 \cite{Fujita2014}.
Indeed, the Fermi arcs seen by STM end precisely on the AF boundary.


\subsubsection{Theoretical models}

The Hubbard model, with a repulsion $U$ between electrons on the same Cu site and an energy $t$ for hopping between sites,
captures many of the experimental properties of cuprates.
On the electron-doped side, the Hubbard model with a moderate repulsion, $U \simeq 6 t$ near optimal doping, accounts for the phase diagram --
with AF order at low $x$ and $d$-wave superconductivity at higher $x$ -- and for the Fermi-surface reconstruction
by AF order, with hot spots on the AF zone boundary \cite{Kyung2004}.
In two dimensions, with decreasing temperature, it accounts for the opening of a 'pseudogap' below the temperature where the AF correlation length
exceeds the thermal de Broglie wavelength of the electrons, {\it i.e.}~when each electron sees its local environment as having AF order \cite{Vilk1997}.
It seems that here there is no fundamental mystery
-- except perhaps the mechanism for $T$-linear resistivity at \xstar~(sec.~6)

Applying the Hubbard model to hole-doped cuprates by increasing $U/t$ leads to a qualitative change.
Indeed, when $U/t$ exceeds a critical value of 6 or so, then a pseudogap phase forms at $T=0$ \cite{Gull2010}.
At $T=0$, it onsets as a transition with decreasing $p$, perhaps first-order \cite{Sordi2012},
but as a crossover with decreasing $T$.
Recent calculations find that the inequality \pstar~$\leq p_{\rm FS}$ holds within the Hubbard model \cite{Wu2017,Braganca2017},
in agreement with experiment \cite{Benhabib2015,Doiron-Leyraud2017}.
The pseudogap leads to a partial loss of DOS, but it is not clear yet what its signature is in the Hall coefficient or the Seebeck coefficient.
This pseudogap comes from short-range AF correlations (spin singlets), not the long-range correlations central to the electron-doped phenomenology.
It is not yet clear what is the Fermi surface inside this pseudogap phase, whether arcs or closed nodal hole pockets containing $p$ carriers (Fig.~\ref{FS}a).
Given that
broken translational symmetry has not been detected so far,
pockets containing $p$ holes would violate the Luttinger rule,
which
can be reconciled
by having
a state with topological order \cite{Sachdev}.

In summary, both the empirical route and the theoretical route so far lead us to a fork in the road:
either hole-doped cuprates are in essence like electron-doped cuprates, but with much shorter AF correlations,
or they have topological order (or some other order?).
In either case, this is a remarkable ground state with no
prior
analog.

\subsubsection{Broken symmetries}

Apart from short-range AF or SDW order that could break translational symmetry over a limited length scale,
we have not discussed broken symmetries so far in connection with the pseudogap phase of hole-doped cuprates.
(CDW order breaks translational symmetry, but it is a separate phase.)
There is substantial (but not yet definitive) experimental evidence that two symmetries are broken below \Tstar:
time-reversal symmetry -- detected as $Q=0$ magnetism via neutrons \cite{Fauque2006} -- and rotational symmetry
-- detected as an extra in-plane anisotropy in the magnetic susceptibility \cite{Sato2017}.
These broken symmetries have been associated with current loop order \cite{Varma2016} and nematic order \cite{Nie2014}, respectively.
The trouble with such orders is that neither can cause a gap to open or the carrier density to drop.
Therefore, they cannot be the driving mechanism for the pseudogap phase, but are perhaps accompanying
 instabilities, a bit like CDW order.
 %
 If so, the question are: What role does current-loop or nematic order play in the pairing?
 In the $T$-linear resistivity? In the mass enhancement above \pstar?


%



\section{STRANGE METAL}

%
%

We call 'strange metal' the region of the phase diagram immediately above \pstar,
extending up to the end of the \Tc~dome (Fig.~\ref{PhaseDiagram}a).
In other words, the metal above \pstar~is a strange metal up until it becomes a normal metal, or Fermi liquid
(sec. 3).
%
At high temperature, this strange metal is characterized by a non-saturating resistivity that exceeds the Ioffe-Mott-Regel limit. At low temperature, it is characterized
by an anomalous $T$ dependence of the resistivity, deviating from the standard $T^2$ behavior.
This is illustrated in Fig.~\ref{NFL}c, where the exponent of the $T$ dependence in $\rho(T)$ for LSCO is mapped as a function of $p$ and $T$ \cite{Cooper2009}.
We see that the exponent is 2.0 at $p = 0.33$, and  it evolves gradually towards 1.0 as $p \to$~\pstar~$\simeq 0.18-0.19$.
The evolution can also be described as a sum of two terms, {\it i.e.} $\rho \sim T+T^2$,
very similar to what is observed in the organic superconductor (TMTSF)$_2$PF$_6$
above its AF QCP \cite{Doiron-Leyraud2009}.
Tl2201 exhibits a similar evolution \cite{Manako1992,Mackenzie1996,Proust2002,Hussey2013},
showing that this is likely to be a generic behavior in hole-doped cuprates.
%
%
ADMR data in Tl2201 have been modelled with two scattering rates:
a $T^2$ rate which is isotropic around the FS,
and an anisotropic $T$-linear rate that is maximal in the antinodal directions \cite{Abdel-Jawad2007}.
The latter term
increases
linearly with \Tc~\cite{Abdel-Jawad2007}.
%
%

Focusing at low temperature, two aspects are striking.
First, the fact that superconductivity emerges in tandem with the deviation from $T^2$ behavior, both starting below the same doping,
in close analogy with (TMTSF)$_2$PF$_6$.
This links $d$-wave pairing with $T$-linear scattering
\cite{Taillefer2010}.
%
Secondly, below a certain doping, $p \simeq 0.27$ in LSCO, $\rho(T)$ becomes perfectly $T$-linear at low $T$.
In Nd-LSCO, this perfect linearity is observed at $p = 0.24$ (Fig.~\ref{NFL}b).
%
%
%
%
It was recently observed in Bi2212 \cite{Legros2018},
and it is also seen in the electron-doped cuprates PCCO \cite{Fournier1998} and LCCO \cite{Sarkar2017,Jin2011},
albeit limited to lower temperatures.
The $T$-linear resistivity as $T \rightarrow 0$ is thus a generic property of cuprates and it is robust against changes in the shape,
topology and multiplicity of the FS.
%

While no compelling explanation for the $T$-linear resistivity as $T \to 0$ has yet been found,
it was observed empirically that the strength of the $T$-linear resistivity for several metals is approximately
given by a scattering rate that has a universal value, namely $\hbar / \tau  = k_{\rm B} T$ \cite{Zaanen2004,Bruin2013}.
This observation suggests that a $T$-linear regime will be observed whenever $1 / \tau$ reaches its Planckian limit, $k_{\rm B} T / \hbar$,
irrespective of the underlying mechanism for inelastic scattering.
Assuming that the connection between $\rho$ and $\tau$ is given by the Drude formula, the linear coefficient of the resistivity $\rho = \rho_0 + A_1 T$  is given (per CuO$_2$ plane) by:
$A_1^* = A_1 / d = (h / 2 e^2)(1/T_F)$, where $T_F = (\pi \hbar^2/k_B) (n d / m^*)$ is the Fermi temperature.
In the overdoped region ($p >$\pstar), the full FS is restored and the carrier density does not vary much with doping.
This implies that $A_1^* \sim$~\mstar.
%
%
Given the effective mass deduced from quantum oscillations or the electronic specific heat, the estimation of $A_1^*$ for both hole-doped and electron-doped cuprates
reveals that the scattering rate responsible for the $T$-linear resistivity has the universal value given by the Planckian limit,
within error bars
\cite{Legros2018}.
This explains why the slope of the $T$-linear resistivity is $\sim 5$ times larger in hole-doped cuprates
({\it i.e.} $A_1^* = 8~\Omega /$K in Nd-LSCO at $p = 0.24$ and Bi2212 at $p = 0.23$)
than in electron-doped cuprates
({\it i.e.} $A_1^* = 1.7~\Omega /$K in PCCO and LCCO at $p = 0.17$),
since the effective mass is $\sim 5$ times higher in the former
\cite{Legros2018}.
%
It also explains why $A_1$ increases in LSCO when going from $p = 0.26$ to $p = 0.21$
\cite{Cooper2009},
since \mstar~rises with decreasing $p$,
as seen from specific heat data
\cite{Legros2018}.
Moreover, a Planckian limit on scattering provides an explanation for the anomalously broad range in doping over which $\rho \sim A_1 T$ is observed in LSCO \cite{Cooper2009}.
As doping decreases below $p = 0.33$,
$A_1^*$ increases steadily until \pstar~$\simeq 0.18-19$, but the scattering rate $1 / \tau$ cannot exceed the Planckian limit,
reached at $p \simeq 0.26$.
So between $p = 0.26$ and $p =$~\pstar, $\rho$(T) is linear and $1 / \tau$ is constant.
However, \mstar~continues to increase until \pstar, so that $A_1^* \sim m^*$ in the range \pstar~$< p < 0.26$.
Understanding the inner workings of the Planckian principle will be a fascinating theoretical challenge. All the more important since in cuprates there is a clear link between $T$-linear scattering and pairing
\cite{Taillefer2010}.

\section{CONCLUSION}
%
We have surveyed the ground state properties of hole-doped cuprates, at $T \to 0$, once superconductivity is removed by the application of a magnetic field.
The central feature
is the critical point \pstar~at which the pseudogap phase onsets.
Two of its key signatures have recently been unveiled.
First,
a drop in carrier density, signalling a transformation of the large Fermi surface above \pstar~into small hole-like
pockets or arcs below \pstar.
Second, a sharp peak in
the electronic specific heat at \pstar, interpreted as
a strong mass enhancement above \pstar~followed
by a gap opening below \pstar.
These signatures are reminiscent of what happens at an antiferromagnetic quantum critical point,
the scenario relevant for electron-doped cuprates.
The remarkable aspect of hole-doped materials is that no long-range order is seen just below \pstar,
raising the possibility of a novel state without broken translational symmetry,
perhaps with topological order.
What does break translational symmetry is CDW order, but only at dopings distinctly below \pstar.
In the CDW phase, the Fermi surface is reconstructed and electron-like,
and, remarkably, its carriers obey Fermi-liquid theory even if ensconced inside the pseudogap phase.
Above \pstar, charge carriers display a fascinating property:
the electrical resistivity shows a perfectly linear temperature dependence as $T \to 0$.
The recent finding that its slope is set by an inelastic scattering rate at the Planckian limit opens
a new perspective on the origin of this archetypal non-Fermi-liquid behavior.
\\

%

\begin{summary}[SUMMARY POINTS]
\begin{enumerate}
\item The organizing principle of electron-doped cuprates is an antiferromagnetic QCP
where long-range AF correlations disappear, at which $T$-linear resistivity is found,
 below which the Fermi surface is reconstructed, and around which $d$-wave superconductivity forms.
\item The thermodynamic signature of the pseudogap critical point \pstar~is a peak in the electronic specific heat at low $T$,
with a $T$log$T$ variation as $T \to 0$ at \pstar.
These are the classic signatures of a QCP
-- but a diverging length scale is still missing...
\item The key transport signature of the pseudogap phase is a drop in the carrier density from $n \simeq 1 + p$ at $p >$~\pstar~to
 $n \simeq p$ at $p <$~\pstar.
\item
The remarkable aspect of the pseudogap ground state is that the Fermi surface is transformed and the carrier density reduced
without long-range order to break the translational symmetry.
A possible explanation is a state with topological order.
\item
The CDW order that forms generically in hole-doped cuprates at $p \simeq 0.12$ produces a Fermi-liquid ground state
inside the pseudogap phase, with a reconstructed Fermi surface that contains a small electron-like pocket.
%
\item
As doping increases beyond $p = 0.12$, the CDW phase weakens and disappears distinctly before the critical doping \pstar~at which
the pseudogap phase ends, thus creating an interval in which the ground state has a pseudogap and a low carrier density without
CDW order.
%
\item The $T$-linear resistivity observed in cuprates at low temperature as $p$ approaches \pstar~from above
is controlled by a universal Planckian limit on the scattering rate.
\item Superconductivity springs from the Fermi-liquid ground state at high doping, and it emerges in tandem with
the inelastic scattering process responsible for the $T$-linear resistivity of the normal state.
Pairing and scattering appear to be linked.
\end{enumerate}
\end{summary}

\begin{issues}[FUTURE ISSUES]
\begin{enumerate}

\item Is the QCP in hole-doped cuprates associated with AF (or SDW) correlations (perhaps short-ranged)?
Or does it differ fundamentally from the QCP in electron-doped cuprates?
%
\item Is SDW order (perhaps short-ranged) generically present in the pseudogap phase of cuprates at $T=0$ when superconductivity (and CDW order) is removed?
\item What is the Fermi surface in the pure pseudogap phase? Arcs or closed nodal hole pockets with $n = p$?
\item
Is there topological order in the pseudogap phase?
How can it be detected?
%
\item Why is CDW order peaked at $p \simeq 1/8$?
\item How does the Planckian limit on inelastic scattering work?
\item Are AF spin fluctuations responsible for $d$-wave pairing? For $T$-linear scattering? For mass enhancement above \pstar?

\end{enumerate}
\end{issues}


\section*{ACKNOWLEDGMENTS}
%
We are thankful for the many stimulating discussions with our colleagues at the workshop
on cuprates in Jouvence, Canada in May 2017, sponsored by the Institut Quantique of Universit\'e de Sherbrooke,
the Canadian Institute for Advanced Research and the EPiQS initiative of the Gordon and Betty Moore Foundation,
namely:
N.P.~Armitage,
W.~Atkinson,
C.~Bourbonnais,
P.~Bourges,
J.~Chang,
A.V.~Chubukov,
J.C.~Davis,
N.~Doiron-Leyraud,
P.~Fournier.
R.L.~Greene,
A.~Georges,
N.E.~Hussey,
M.-H.~Julien ,
A.~Kaminski,
S.A.~Kivelson,
 G.~Kotliar,
B.~Keimer,
D.-H.~Lee,
A.J.~Millis,
B.J.~Ramshaw,
M.~Randeria,
 T.M.~Rice,
 S.~Sachdev,
 D.J.~Scalapino,
 J.~Schmalian,
 S.E.~Sebastian,
 D.~S\'en\'echal,
 G.~Sordi,
J.L.~Tallon,
 J.~Tranquada,
 A.-M.S.~Tremblay,
 D.~van der Marel.
We also thank our
numerous collaborators on high-field experiments around the world,
with whom it was a privilege and a pleasure to discover and explore the ground-state
properties of non-superconducting cuprates.
%
We acknowledge the kind support and hospitality
of the Institut Quantique in Sherbrooke (C.P.)
and
the Labex NEXT and LNCMI in Toulouse (L.T.)
while this article was written.

\end{document}